\RequirePackage{fix-cm}
\documentclass[smallextended]{svjour3}

\usepackage[
  paperwidth=155mm,
  paperheight=235mm,
  textwidth=119mm,
  textheight=194mm,
  hcentering,
  vcentering
]{geometry}

\usepackage{marvosym}

\journalname{Operations Research Forum}

\usepackage[T1]{fontenc}
\usepackage[utf8]{inputenc}
\usepackage{amsmath,amssymb,bm}
\usepackage{array,booktabs,makecell}
\usepackage{graphicx}
\usepackage{float}
\usepackage{pdflscape}
\usepackage{enumitem}
\usepackage[numbers,sort&compress]{natbib}
\usepackage{url}
\usepackage[
  colorlinks=true,
  linkcolor=blue,
  urlcolor=blue,
  citecolor=blue,
  anchorcolor=blue
]{hyperref}

\newcolumntype{P}[1]{>{\centering\arraybackslash}p{#1}}
\newcommand{\xG}{\mathrm{xG}}
\newcommand{\xA}{\mathrm{xA}}
\newcommand{\xGI}{\mathrm{xGI}}
\newcommand{\xGC}{\mathrm{xGC}}
\begin{document}
\raggedbottom

\title{A data-driven framework for team selection in Fantasy Premier League}

\author{Danial Ramezani \and Tai Dinh\textsuperscript{\Letter}}

\institute{
  Danial Ramezani
  \at HEC Li\`ege School of Management, University of Li\`ege,
  Rue Louvrex 14, 4000, Li\`ege, Belgium\\
  \email{danial.ramezani@uliege.be}
  \and
  \textsuperscript{\Letter}Tai Dinh
  \at The Kyoto College of Graduate Studies for Informatics, Japan\\
  \email{t\_dinh@kcg.ac.jp}
}
\date{}
\authorrunning{Danial Ramezani and Tai Dinh}
\maketitle

\begin{abstract}
Fantasy Premier League (FPL) requires managers to allocate a \pounds 100m budget across a 15-player squad and choose a legal starting eleven, captain, and bench under uncertainty about future player points. We integrate player-level forecasts with mixed-integer linear programming (MILP) and compare a two-stage heuristic, which selects starters before reserves, with a joint 15-player model that allocates expenditure endogenously between starters and the bench. An epsilon-constraint frontier characterizes the predicted starter--bench trade-off. Using Gameweeks (GWs) 1--26 of the 2023/24 season for model construction and GWs 27--38 for evaluation, the static policy retains the initial lineup and bench order, whereas the fixed-squad sequential policy updates only the starting eleven and bench order; squad membership and captaincy remain fixed. Under autoregressive integrated moving average (ARIMA) forecasts with order $(1,0,0)$, joint selection scores 743 static and 741 sequential points, compared with 643 under both policies for the heuristic \pounds 83.5m/\pounds 16.5m split. Under weighted averaging, the joint and actual-remainder two-stage designs coincide, scoring 721 static and 728 sequential points. Additional analyses examine robust objectives, regularization, coefficient stability, and player availability. The framework supports transparent one-time squad construction and fixed-squad lineup management rather than season-long transfer planning.
\end{abstract}


\section{Introduction}

Fantasy sports are widely used decision games in which participants transform forecasts of real-player performance into resource-constrained rosters. The official Fantasy Premier League (FPL) competition is a particularly transparent setting: a manager constructs a 15-player squad under a \pounds 100.0m budget and makes lineup decisions over 38 gameweeks. The problem combines uncertain rewards with cardinality, positional, formation, captaincy, and club-quota constraints. It is therefore both a practical sports-analytics application and a compact resource-allocation problem.

An FPL squad contains two goalkeepers (GK), five defenders (DEF), five midfielders (MID), and three forwards (FWD), with at most three players from one Premier League club. Before each deadline, the manager chooses 11 starters in a legal formation: exactly one goalkeeper, three to five defenders, two to five midfielders, and one to three forwards. Only players who appear score for the planned XI; the captain's score is doubled. If a starter records no minutes, official-style automatic substitution uses the ordered bench while preserving a legal formation, and a designated vice-captain replaces a non-appearing captain. Transfers beyond a manager's available free-transfer allowance normally incur point deductions, and season-limited chips modify selected rules. Our experiment deliberately studies a narrower decision: a new squad is constructed once at GW27 and then held fixed through GW38. No external transfers, transfer-hit deductions, or chips occur after initialization.

Within this fixed-squad setting, we distinguish two lineup policies. The \emph{static policy} keeps the planned GW27 starting XI and bench order unchanged throughout the evaluation horizon, apart from formation-valid automatic substitutions. The \emph{fixed-squad sequential policy} updates player forecasts before each gameweek using only earlier observations and may reassign the starting XI and bench order within the same 15-player squad. Squad membership, captain, and vice-captain remain fixed under both policies. Thus, ``sequential'' refers to tactical lineup updating rather than repeated squad construction, transfers, or dynamic captaincy.

Prior FPL research has often concentrated on predicting player points, leaving the downstream constrained decision either implicit or only partly modeled. Conversely, an optimizer can be operationally misleading if its cost vector is unstable or if its formulation imposes an arbitrary budget allocation. We therefore study the full prediction--optimization link and distinguish forecast quality, roster design, lineup policy, and realized evaluation.

The paper makes three scientific contributions.
\begin{itemize}[leftmargin=1.5em]
  \item We develop and compare two MILP-based roster designs. A transparent two-stage heuristic optimizes the starting XI and captain before selecting the bench, using either a fixed starter--bench allocation or an actual-remainder budget rule. A joint 15-player MILP instead selects the squad, starting XI, and captain simultaneously, making starter and bench expenditure endogenous; an epsilon-constraint analysis exposes their predicted-value trade-off without an arbitrary weighted-sum coefficient.
  \item We examine how feature redundancy and regularization affect the stability of player forecasts and the resulting roster decisions. Full, reduced, and parsimonious Ridge specifications are compared with LASSO using leakage-safe rolling-origin tuning and 500 player-cluster bootstrap replications.
  \item We conduct a controlled internal evaluation of static and fixed-squad
    sequential lineup policies under a common forecasting and roster-initialization
    protocol, including automatic substitutions, robust objectives, budget-policy
    sensitivity, and availability-aware analysis. Selected results are also compared
    descriptively with externally supplied benchmarks.
\end{itemize}
The public repository provides the data, scripts, configuration records, and output manifests used to reproduce the analysis and verify the reported results.

The remainder of the paper is organized as follows. Section~\ref{related_work} reviews the relevant operations-research and fantasy-sports literature. Section~\ref{sec:prelim} formulates the FPL decision models. Sections~\ref{sec:framework} and~\ref{sec:Dataset} describe the forecasting framework, data, and experimental design, while Section~\ref{sec:results} presents and discusses the consolidated results. Finally, Sections~\ref{sec:limitations} and~\ref{sec:conclusion} discuss the study's limitations and present the conclusions and directions for future research. Supplementary Sections~S1--S7 provide detailed forecasting equations, exploratory and feature diagnostics, full method screens, budget and lineup-policy sensitivity analyses, descriptive comparisons, numerical tables, and computational reproducibility information.

% ------------------------------------------------------
\section{Related work}\label{related_work}

Operations research (OR) in sport spans scheduling, forecasting, tactics,
and selection decisions, as reviewed by Wright~\cite{wright2009ORsport}.
From an OR perspective, FPL roster construction is a
cardinality-constrained resource-allocation problem: a finite set of players
must cover positional requirements while satisfying club and budget
restrictions, and the objective coefficients are estimated from uncertain
player returns. Related sports-selection models include integer programming
for selecting a Twenty20 cricket team~\cite{sharp2011cricket},
multiobjective optimization for football team-member selection under budget
restrictions~\cite{zhao2021multiobjective}, and data-driven football team
formation formulated as a maximum-weight matching
problem~\cite{nouraie2023intelligent}. Ati et al.~\cite{ati2024using}
provide a systematic review of multi-criteria decision-making and
machine-learning methods for football player selection and performance
prediction.

The same mathematical structure has clear analogues outside sport.
Workforce-rostering models impose qualification and coverage
constraints~\cite{ernst2004staff}; portfolio-selection models allocate scarce
capital while balancing expected return and
risk~\cite{markowitz1952portfolio,mansini2014portfolio}; and robust
optimization provides tractable protection against uncertainty in objective
coefficients and other model
parameters~\cite{bertsimas2004price,gabrel2014robust}.
Epsilon-constraint methods are particularly relevant to the starter--bench
trade-off because they identify nondominated alternatives without requiring
a single fixed weighted-sum coefficient~\cite{mavrotas2009epsilon}.

Within fantasy sports, Matthews et al.~\cite{matthews2012competing}
formulate fantasy-football management as a partially observable sequential
decision problem. Becker and Sun~\cite{becker2016analytical} use
mixed-integer programming for drafting and weekly lineup decisions in
fantasy football based on the National Football League (NFL).
Gupta~\cite{gupta2019time} combines a recurrent neural network
(RNN)--ARIMA forecaster with linear FPL selection, while Venter and van
Vuuren~\cite{venter2024optimisation} and Aribo~\cite{aribo2024machine}
integrate prediction with optimization or metaheuristics. Beal et
al.~\cite{beal2020dfs} likewise demonstrate the value of connecting
predictive models to constrained team formation rather than ranking players
independently.

A parallel literature focuses on the predictive layer. Bangdiwala et
al.~\cite{bangdiwala2022using} compare machine-learning regressors for FPL
points; Pokharel et al.~\cite{pokharel2022fantasy} study selection and
transfers using boosted trees; and Lombu et al.~\cite{lombu2024predicting}
compare convolutional and recurrent sequence models. Bhatt et
al.~\cite{bhatt2019should} investigate crowd signals for captaincy. These
studies motivate alternative cost estimators but do not directly examine how
redundancy in those estimates affects the stability of the cost vector or
the resulting constrained roster.

The closest external FPL benchmark in our evaluation is the integrated
prediction-and-MILP study of Santoro~\cite{santoro2025FPL}, whose four
strategies span greedy construction, random lineup choice, score-ranked MIP
lineups, and online score updating. They are relevant integrated baselines
rather than an exhaustive definition of the state of the art: public FPL
studies differ in seasons, data, transfer assumptions, and evaluation
protocols. Our contribution is consequently methodological and
diagnostic---a unified, auditable comparison of cost estimation, roster
construction, robustness, and fixed-squad lineup decisions under one dataset
and one evaluation protocol.

% -------------------------------------------------------------

\section{FPL decision models}\label{sec:prelim}

Let $j=1,\ldots,n$ index candidate players. Player $j$ has price $v_j$, predicted reward $c_j$, position $k(j)$, and club $t(j)$. Let $\mathcal P_k$ and $\mathcal T_t$ denote the corresponding position and club sets. The full-squad budget is $B=100$ (million pounds), the official squad quotas are $(R_{\mathrm{GK}},R_{\mathrm{DEF}},R_{\mathrm{MID}},R_{\mathrm{FWD}})=(2,5,5,3)$, and the legal starting-XI bounds are $(L_k,U_k)=(1,1),(3,5),(2,5),(1,3)$.

\subsection{Two-stage starting-XI and bench optimization}

In the first stage of the two-stage heuristic, binary $x_j$ and $y_j$ indicate that player $j$ starts and is captain, respectively. For a starting-XI cap $b$, the model is
\begin{equation}\label{eq:two_stage_xi}
\begin{aligned}
\max_{x,y}\quad & \sum_j c_jx_j+\sum_jc_jy_j \\
\text{s.t.}\quad
&\sum_jx_j=11,\quad \sum_jv_jx_j\le b,\quad \sum_jy_j=1,\quad y_j\le x_j &&\forall j,\\
&L_k\le\sum_{j\in\mathcal P_k}x_j\le U_k &&\forall k,\\
&\sum_{j\in\mathcal T_t}x_j\le3 &&\forall t,\\
&x_j,y_j\in\{0,1\} &&\forall j.
\end{aligned}
\end{equation}
The second stage selects binary bench variables $z_j$ after the XI has been fixed. If $r_k=R_k-\sum_{j\in\mathcal P_k}x_j$, $s_t=\sum_{j\in\mathcal T_t}x_j$, and $S=\{j:x_j=1\}$, then
\begin{equation}\label{eq:two_stage_bench}
\begin{aligned}
\max_z\quad &\sum_jc_jz_j\\
\text{s.t.}\quad
&\sum_{j\in\mathcal P_k}z_j=r_k &&\forall k,\\
&\sum_jv_jz_j\le b^{\mathrm B},\quad
  \sum_{j\in\mathcal T_t}z_j+s_t\le3 &&\forall t,\\
&z_j=0 &&\forall j\in S,\qquad z_j\in\{0,1\} &&\forall j.
\end{aligned}
\end{equation}
The \emph{fixed-allocation} rule sets $b=83.5$ and $b^{\mathrm B}=B-b=16.5$. This is a reproducible heuristic based on reserving approximately the minimum feasible bench expenditure, not an OR-derived optimum. The \emph{actual-remainder} variant instead sets $b^{\mathrm B}=B-\sum_jv_jx_j$, so unspent XI allowance is available to the bench.

\subsection{Joint 15-player optimization and Pareto analysis}\label{sec:joint_formulation}

To remove the exogenous split, let binary $q_j$, $u_j$, and $y_j$ indicate squad membership, starting-XI membership, and captaincy. Define starter and bench value as
\begin{equation}\label{eq:joint_objectives}
F_{\mathrm{XI}}(u,y)=\sum_jc_ju_j+\sum_jc_jy_j,
\qquad
F_{\mathrm B}(q,u)=\sum_jc_j(q_j-u_j).
\end{equation}
The joint feasible set $\Omega$ is
\begin{equation}\label{eq:joint_feasible_set}
\begin{aligned}
\Omega=\bigg\{(q,u,y):\;&
\sum_j q_j=15,\quad
\sum_j u_j=11,\quad
\sum_j y_j=1,\\
&
u_j\leq q_j,\quad
y_j\leq u_j,\quad
q_j,u_j,y_j\in\{0,1\}
&& \forall j,\\
&
\sum_j v_jq_j\leq B,\\
&
\sum_{j\in P_k}q_j=R_k,\quad
L_k\leq\sum_{j\in P_k}u_j\leq U_k
&& \forall k,\\
&
\sum_{j\in T_t}q_j\leq3
&& \forall t
\bigg\}.
\end{aligned}
\end{equation}
Thus $b^{\mathrm{XI}}=\sum_jv_ju_j$ and $b^{\mathrm B}=\sum_jv_j(q_j-u_j)$ are endogenous and satisfy $b^{\mathrm{XI}}+b^{\mathrm B}\le B$.

We first obtain $F_{\mathrm{XI}}^*=\max_{(q,u,y)\in\Omega}F_{\mathrm{XI}}$. We evaluate the prespecified proportional XI-loss grid
\[
\mathcal D=\{0,0.005,0.01,0.02,0.05,0.10\}.
\]
For each $\delta\in\mathcal D$, we solve
\begin{equation}\label{eq:pareto}
\max_{(q,u,y)\in\Omega}F_{\mathrm B}(q,u)
\quad\text{s.t.}\quad
F_{\mathrm{XI}}(u,y)\ge(1-\delta)F_{\mathrm{XI}}^*.
\end{equation}
A lexicographic refinement maximizes $F_{\mathrm{XI}}$ among solutions with the same optimal bench value. The displayed compromise minimizes Euclidean distance to the forecast-space utopia point after min--max normalization of both objectives. Neither $\delta$ nor $b$ is selected from realized GW27--38 points.

\subsection{Robust and availability-aware coefficients}\label{sec:robust}

For the box-robust XI model, predicted rewards lie in a box $\mathcal U=\{c:c_j\in[\bar c_j-d_j,\bar c_j+d_j]\}$ and the optimizer maximizes the worst-case XI-plus-captain reward. The resulting objective is equivalent to using the lower coefficient $\bar c_j-d_j$ for each selected contribution. This isolates sensitivity to overestimated player value while preserving the linear integer model; it is related to the broader robustness--performance trade-off in Bertsimas and Sim~\citep{bertsimas2004price}.

We additionally examine future starting uncertainty. If $a_j$ is the number of GWs among 1--26 in which player $j$ starts, missing player--GW rows count as zero, and a Beta$(1,1)$ prior gives $p_j=(a_j+1)/(26+2)$. The expected-availability and uncertainty-penalized coefficients are
\begin{equation}\label{eq:availability}
c_j^{\mathrm{EA}}=p_jc_j,
\qquad
c_j^{\mathrm{AR}}=p_jc_j-\Gamma|c_j|\sqrt{p_j(1-p_j)},
\qquad \Gamma=0.5.
\end{equation}

The second coefficient discourages players whose estimated starting probability is both low and uncertain. This is a sensitivity design, not a claim that historical starting frequency is a complete injury or rotation model.

\clearpage
\newgeometry{left=2mm,right=2mm,top=2mm,bottom=2mm}

\begin{landscape}
\begin{figure}[p]
    \centering
    \includegraphics[
        width=0.77\linewidth
    ]{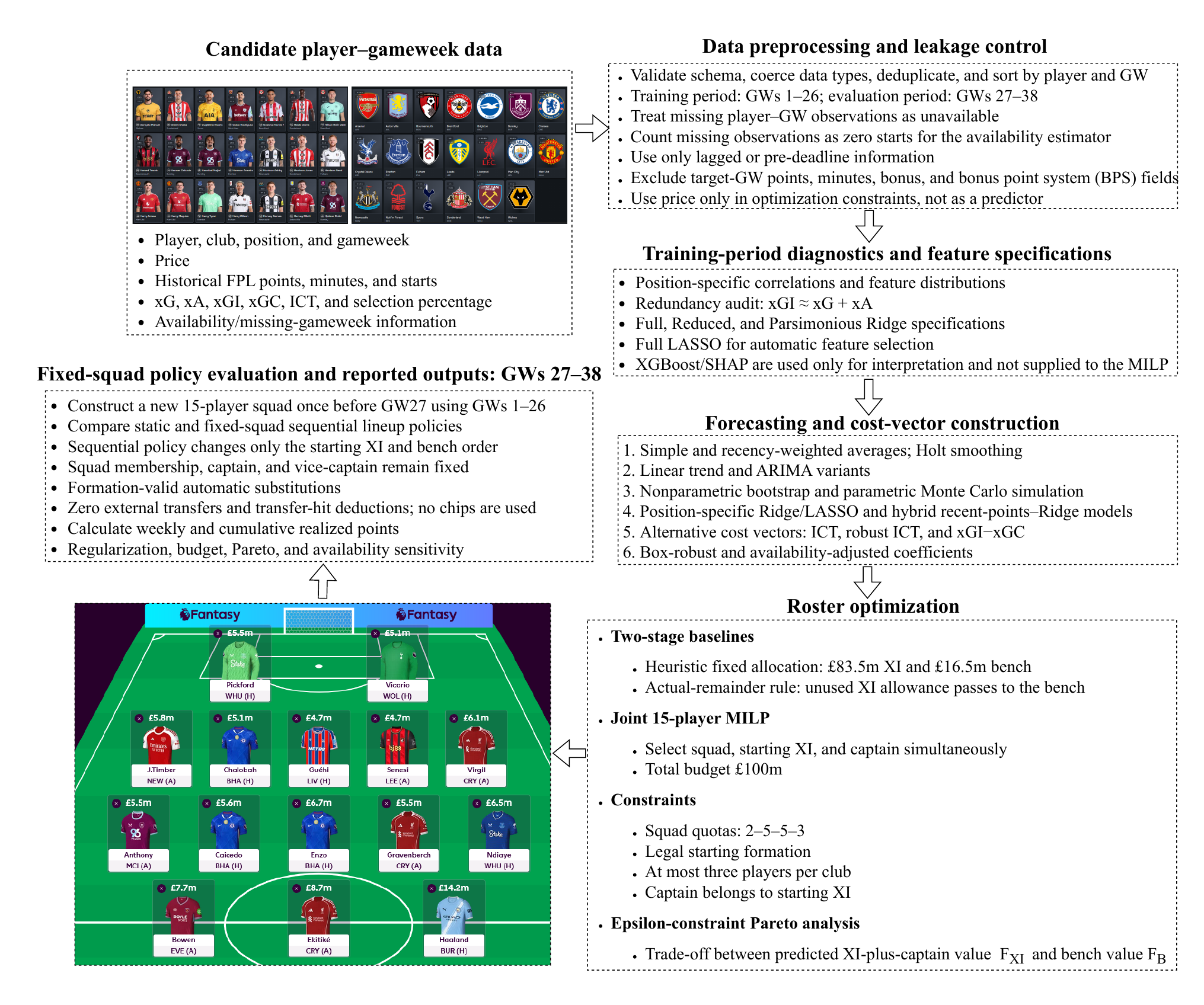}
    \vspace{-0.4cm}
    \caption{End-to-end workflow for data preprocessing, forecasting, roster optimization, and evaluation.}
    \label{fig:framework}
\end{figure}
\end{landscape}

\restoregeometry

\section{Forecasting framework}\label{sec:framework}

Figure~\ref{fig:framework} summarizes the workflow. Each estimator maps player history available before the decision deadline to a cost coefficient $c_j$. The screen includes simple and linearly recency-weighted averages, Holt trend smoothing, nonparametric bootstrap and parametric Monte Carlo simulation, linear time trend, and nine low-order ARIMA specifications. We also evaluate robust versions, alternative influence--creativity--threat index (ICT) and expected-involvement objectives, and hybrid scores. Standard definitions and equations for each estimator are collected in Supplementary Sec.~S1 to keep the main methodological discussion focused on the decision model.

The feature-based hybrid fits a separate standardized regression by position. Its seven candidate inputs are expected goals ($\xG$), expected assists ($\xA$), expected goal involvements ($\xGI$), expected goals conceded ($\xGC$), ICT, selection percentage, and starts. The regression prediction receives weight $2/3$ and the recent-points component receives weight $1/3$. Player price is used only in the budget constraint, not as a predictive input. A diagnostic XGBoost analysis using SHapley Additive exPlanations (SHAP) is reported only for interpretation and does not enter the optimizer.

\begin{figure*}[!htb]
  \centering
  \includegraphics[width=\textwidth]{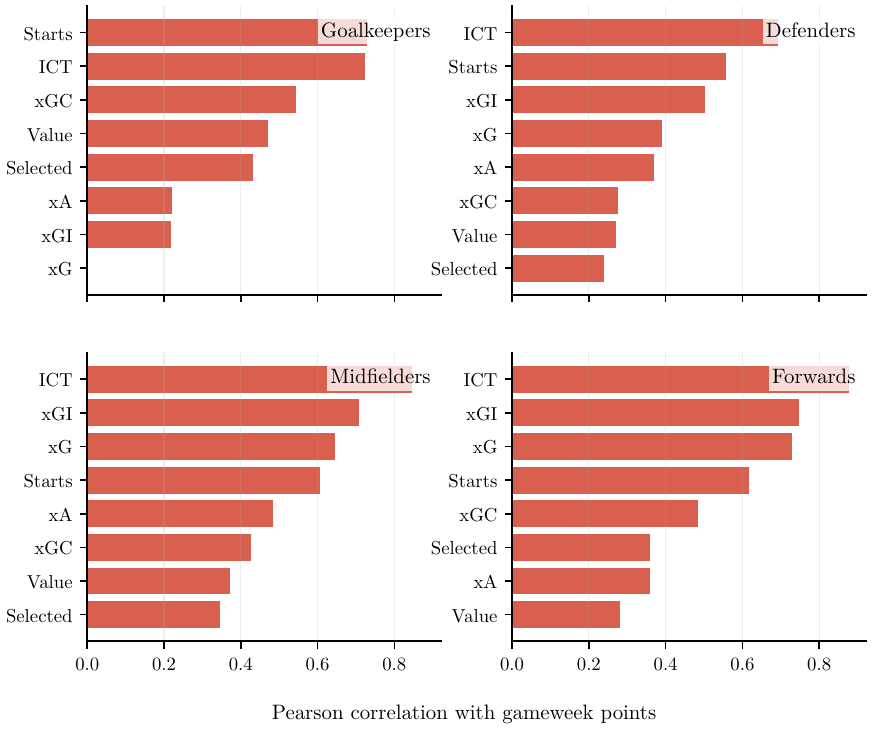}
  \caption{Position-specific Pearson correlations between candidate player attributes and total points in the training period. Price (\emph{Value}) is shown for diagnostic comparison but is not used as a forecasting predictor. Starts and the Influence, Creativity and Threat (ICT) index show relatively strong correlations across positions, while attacking metrics have their strongest associations for midfielders and forwards.}
  \label{fig:position_correlations}
\end{figure*}

Figure~\ref{fig:position_correlations} provides the position-level motivation for the feature-based models. The different correlation profiles support estimating separate regressions by position, while the similar behavior of the attacking summaries motivates the formal redundancy and regularization analysis below.

Ridge regression is well suited to correlated predictors because $L_2$ shrinkage stabilizes coefficients while retaining shared signals~\citep{hoerl1970ridge}. Because $\xGI$, $\xG$, $\xA$, and ICT overlap, we assess whether simpler feature sets or automatic feature selection improve stability without sacrificing predictive performance. We compare: (i) Full Ridge with all seven features; (ii) Reduced Ridge, which removes $\xG$ and $\xA$ because $\xGI=\xG+\xA$; (iii) Parsimonious Ridge, which additionally removes ICT and retains $\xGI$ as the sole attacking feature alongside $\xGC$, selection percentage, and starts; and (iv) Full LASSO, which offers automatic $L_1$ feature selection~\citep{tibshirani1996lasso}. Penalties are chosen separately by position from rolling-origin training folds; LASSO uses the one-standard-error rule to favor sparsity.

\section{Data and experimental design}\label{sec:Dataset}

We use the public 2023/24 player--gameweek panel from the Fantasy-Premier-League repository.\footnote{\url{https://github.com/vaastav/Fantasy-Premier-League}} The exact input snapshot used in this study is archived with the code in our project repository.\footnote{\url{https://github.com/taiduydinh/fplorf}} Prices are converted to millions of pounds, and player, club, and position fields define the optimization constraints. After schema checks and preprocessing, the roster-design run contains 19,327 training rows from GWs 1--26, 10,398 evaluation rows from GWs 27--38, and 832 candidate players. Model-specific complete-case counts can be lower.

The outcome is official gameweek total points. Same-gameweek outcomes---including goals, assists, clean sheets, cards, saves, bonus, and bonus-point-system fields---are never used to predict that gameweek. History-only estimators use only prior total-points sequences. Feature regressions use lagged, pre-deadline aggregates. Missing player--GW observations are treated as non-availability for evaluation and as zero starts in the availability estimator; target-week minutes are never a predictor.

\subsubsection*{Controlled roster designs}
The comparison uses Simple and Weighted averages and ARIMA orders $(0,0,1)$, $(1,0,0)$, and $(1,0,1)$. The fixed and actual-remainder rules use the prespecified cap grid
\[
b\in\{55,60,65,70,75,80,83.5\},
\]
and the joint model uses the six $\delta$ values stated in Sec.~\ref{sec:joint_formulation}. The main table and corresponding figures highlight Weighted average (a recency-weighted historical baseline) and ARIMA(1,0,0) (a first-order autoregressive baseline) as complementary illustrations of the budget-allocation mechanisms, not as the highest-scoring forecasters. For Weighted average, allowing unused XI allowance to reach the bench makes the actual-remainder and joint $\delta=0$ designs coincide; for ARIMA(1,0,0), the joint XI costs \pounds 83.7m and therefore cannot be selected under the heuristic \pounds 83.5m XI cap. The complete five-method results are reported in Supplementary Table~S2 and the CSV outputs; ARIMA(0,0,1) achieves the highest realized total among the five joint XI-optimal designs (765 points under both policies). Each MILP was configured with a requested relative optimality gap of zero and a 120-second time limit. An instrumented full-pipeline reproduction logged 5,639 individual MILP solves, all with solver-reported optimal termination and none terminated by the time limit; Supplementary Sec.~S7 reports the termination counts and achieved-gap diagnostics.

\subsubsection*{Static and fixed-squad sequential policies}
Both policies construct a new 15-player squad at GW27 from GWs 1--26; neither inherits an existing GW26 squad. The static policy retains the planned XI and bench order across the evaluation horizon, subject to automatic substitutions. The sequential policy updates player forecasts before each target GW using only earlier observations and may change the starting XI and bench order within the selected GW27 squad. Before the lineup decision for target GW $t\in\{27,\ldots,38\}$, player-point histories and, where applicable, lagged feature aggregates use only observations from GWs earlier than $t$ (GWs 1--26 for GW27); squad-selection prices, teams, and positions remain those observed before GW27, and target-GW points, minutes, and other match outcomes are excluded. Squad membership, captain, and vice-captain remain fixed through GW38. Hence $S_t=S_{27}$ and $|S_t\setminus S_{t-1}|=0$ for $t=28,\ldots,38$: the experiment makes zero external transfers, pays zero transfer-hit points, and does not dynamically change captaincy. All 200 roster-design policy-audit rows and all eight regularization-policy rows pass these invariance checks.

Realized totals include formation-valid automatic substitutions when a planned starter records no minutes. The highest-forecast non-captain starter at initialization is designated vice-captain and receives the additional captain contribution if the captain does not appear. Budgets, $\delta$, regularization penalties, and forecasting specifications are never chosen using realized GW27--38 outcomes.

\subsubsection*{Regularization and availability experiments}
The four regression specifications use rolling-origin development folds ending before GW27. Stability is estimated with 500 position-stratified player-cluster bootstrap samples that preserve each sampled player's complete training history. We report test mean absolute error (MAE), root mean squared error (RMSE), Spearman rank correlation, bootstrap cost-vector correlation and normalized RMSE, squad and XI Jaccard similarity, captain stability, and realized totals. The availability analysis applies Eq.~\ref{eq:availability} to each of the five roster-design forecasts and compares it with the unadjusted joint $\delta=0$ roster.

\paragraph{Computational reproducibility.}
The complete analysis can be reproduced from the raw 2023/24
player--gameweek panel using the public four-module Python pipeline available
in the project repository.\footnote{\url{https://github.com/taiduydinh/fplorf}}
With the fixed seed 20260824, the pipeline regenerates all 94 internal weekly
series, roster decisions, tables, analytical figures, and policy-audit records
without loading previously stored internal results. The four
Santoro~\cite{santoro2025FPL} benchmark series are the only external numerical
inputs and are identified separately in the generated provenance records.
Detailed installation and execution commands, minimum software requirements,
the realized computational environment, input and output checksums, and
instructions for local and server execution are provided in the Supplementary
Material and repository README.

% ------------------------------------------------------------------------------

\section{Results and discussion}\label{sec:results}

\subsection{Feature redundancy, Ridge parsimony, and LASSO}\label{sec:regularization_results}

The data audit confirms a deterministic redundancy: across 19,086 complete training rows, $\xGI=\xG+\xA$ exactly in 99.906\% of rows and to the stored rounding precision in 100\%, with maximum absolute discrepancy 0.01. The correlation between $\xGI$ and $\xG$ is 0.907 and that between $\xGI$ and ICT is 0.803. Including all four attacking summaries is therefore not necessary merely to preserve their shared signal.

\begin{figure*}[!htb]
  \centering
  \includegraphics[width=\textwidth]{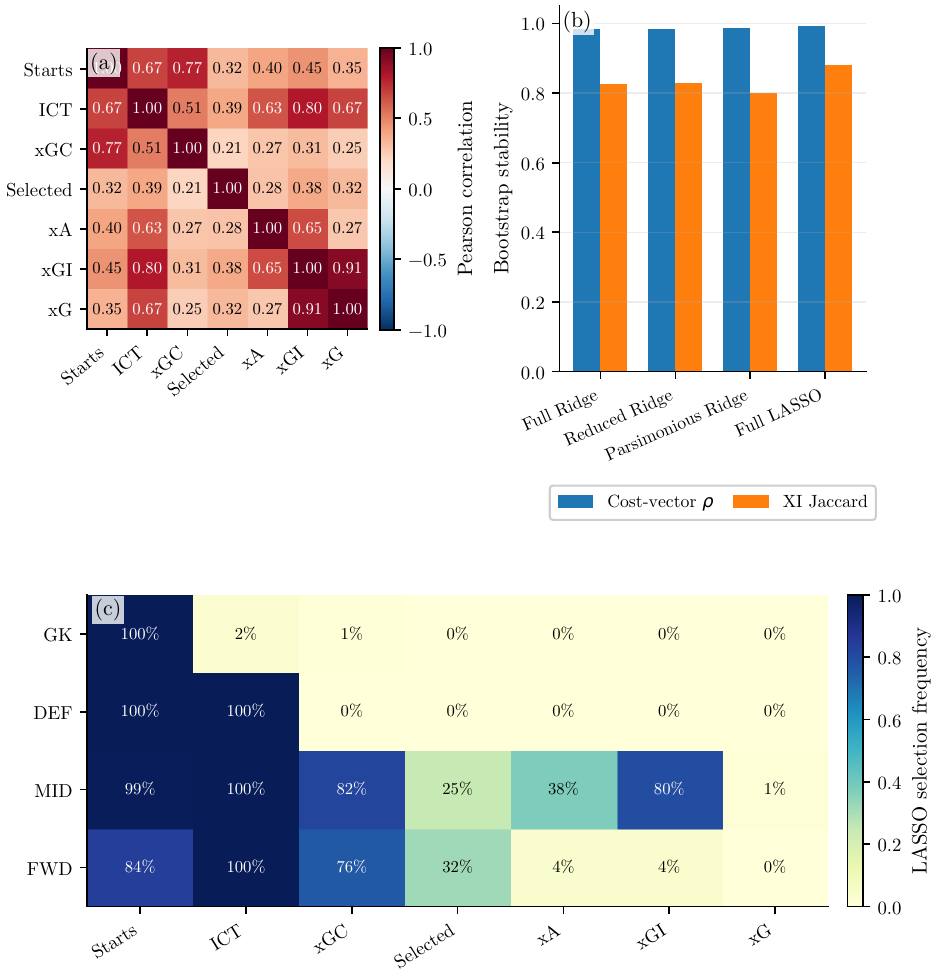}
 \caption{Feature redundancy and regularization evidence. The panels summarize training-period feature correlations, bootstrap cost-vector and starting-XI stability, and position-specific LASSO selection frequencies.}
  \label{fig:regularization}
\end{figure*}

\begin{table}[!htb]
\centering
\caption{Predictive, cost-vector, and decision stability for the four
regression specifications. Test $\rho$ is the test-period Spearman
correlation; Cost $\rho$ is the mean bootstrap Spearman correlation with the
reference cost vector; Norm.\ cost RMSE is the normalized bootstrap
cost-vector RMSE; Squad J.\ and XI J.\ are mean bootstrap Jaccard
similarities; Captain stab.\ is the proportion of bootstrap solutions
retaining the reference captain; and Static and Sequential report realized
points under the two lineup policies.}
\label{tab:regularization}

\setlength{\tabcolsep}{4pt}

\begin{tabular}{@{}lrrrrrr@{}}
\toprule
\multicolumn{7}{c}{\textbf{(a) Predictive performance and cost-vector stability}}\\
\cmidrule(lr){1-7}
Model
& Features
& MAE
& RMSE
& Test $\rho$
& Cost $\rho$
& \shortstack{Norm.\ cost\\RMSE} \\
\midrule
Full Ridge
& 7 & 0.656 & 1.090 & 0.802 & 0.985 & 0.055 \\
Reduced Ridge
& 5 & 0.657 & 1.093 & 0.801 & 0.985 & 0.051 \\
Parsimonious Ridge
& 4 & 0.658 & 1.080 & 0.799 & 0.986 & 0.053 \\
Full LASSO
& 7 & 0.709 & 1.062 & 0.797 & 0.992 & 0.053 \\
\bottomrule
\end{tabular}

\vspace{0.9em}

\begin{tabular}{@{}lrrrrr@{}}
\toprule
\multicolumn{6}{c}{\textbf{(b) Decision stability and realized performance}}\\
\cmidrule(lr){1-6}
Model
& Squad J.
& XI J.
& \shortstack{Captain\\stab.}
& Static
& Sequential \\
\midrule
Full Ridge
& 0.645 & 0.825 & 0.978 & 640 & 639 \\
Reduced Ridge
& 0.660 & 0.829 & 0.956 & 640 & 639 \\
Parsimonious Ridge
& 0.658 & 0.802 & 0.980 & 687 & 689 \\
Full LASSO
& 0.917 & 0.882 & 0.862 & 556 & 558 \\
\bottomrule
\end{tabular}
\end{table}

Figure~\ref{fig:regularization} and Table~\ref{tab:regularization} show that reducing the Ridge input set does not materially damage predictive ranking or cost-vector stability. Reduced Ridge slightly improves normalized cost RMSE and decision Jaccard similarity, while Parsimonious Ridge has the lowest Ridge RMSE and higher realized totals in this one horizon. These differences are not uniform across every metric, so we do not claim that one reduced Ridge specification universally dominates Full Ridge.

LASSO answers a different question. It produces the highest bootstrap cost correlation (0.992) and squad stability (0.917) and the lowest overall test RMSE (1.062), but its MAE is worse and its realized totals are lower. Its sparse position-specific solutions consistently retain starts and ICT for defenders, ICT for attackers, and starts for goalkeepers; $\xG$ is almost always set to zero, while $\xGI$ is retained most often for midfielders. Thus LASSO is useful when sparse and stable selection is the priority, but automatic feature selection does not guarantee superior downstream points. Full Ridge serves as the principal hybrid estimator because its MAE is competitive and its $L_2$ penalty accommodates correlated signals; the reduced Ridge specifications and LASSO provide validated alternatives.

\subsection{Controlled predictive and lineup performance}

\begin{figure*}[!htb]
  \centering
  \includegraphics[width=\textwidth]{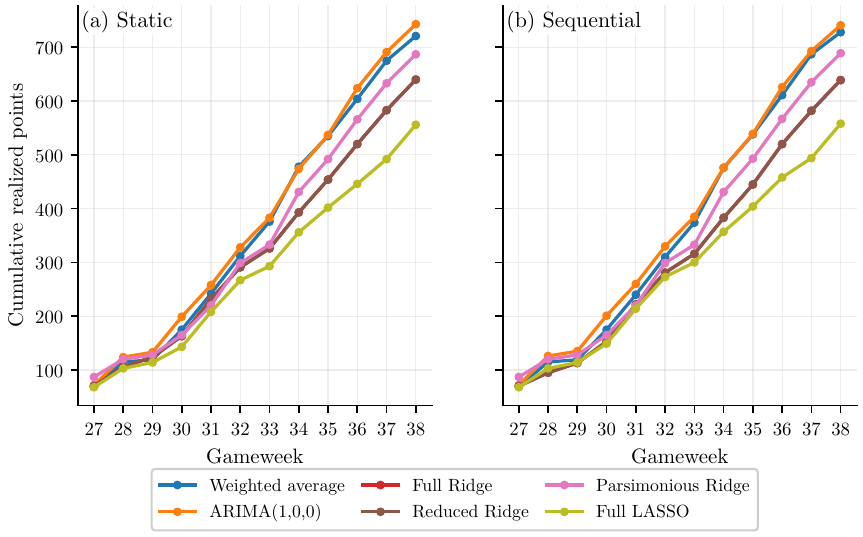}
  \caption{Realized GW27--38 totals for two representative joint $\delta=0$ forecasters---Weighted average and ARIMA(1,0,0)---and the four regularization specifications under static and fixed-squad sequential policies. Complete five-forecaster roster-design results are reported in Supplementary Table~S2.}
  \label{fig:primary_performance}
\end{figure*}

Figure~\ref{fig:primary_performance} presents the two representative roster-design forecasters alongside the four regularization specifications. The complete five-forecaster comparison in Supplementary Table~S2 shows that ARIMA(0,0,1) records 765 points under both policies, ARIMA(1,0,1) records 737 static and 761 sequential points, ARIMA(1,0,0) records 743 and 741, and Weighted average records 721 and 728. These results establish low-order ARIMA and recency weighting as strong baselines in this sample, but they do not identify a universal winner. Within this comparison, ARIMA(0,0,1) has the highest realized total, while ARIMA(1,0,1) has the largest gain from sequential lineup updating. The complete 94-series forecast and objective screen is reported in Supplementary Figs.~S5--S24 and the accompanying CSV table; the test-period rankings in Figs.~S23--S24 are explicitly post-evaluation descriptions, not model-selection evidence.

\begin{figure*}[!htb]
  \centering
  \includegraphics[width=\textwidth]{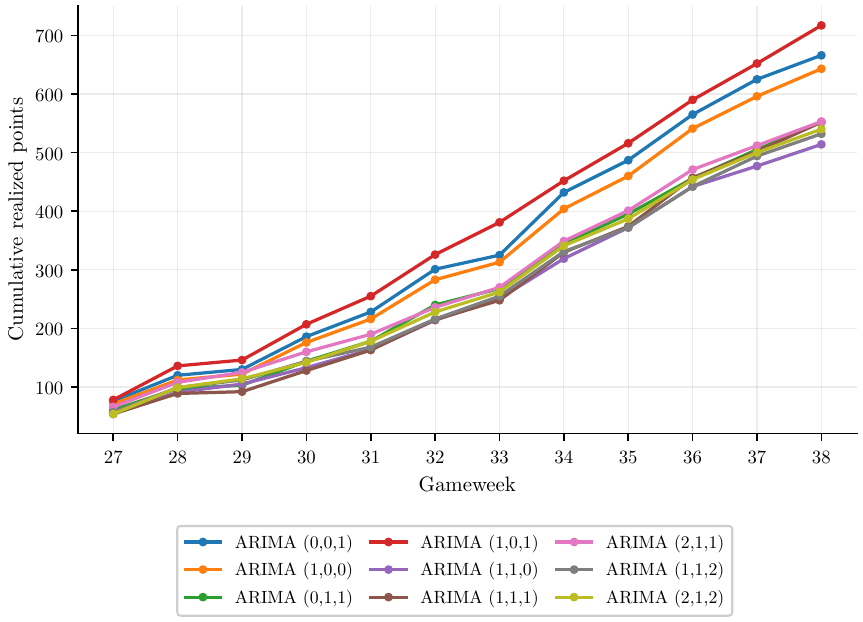}
  \caption{Cumulative realized performance of the nine ARIMA specifications in the broad static screen. Low-order specifications are competitive in this sample, whereas increasing model order does not consistently improve the downstream decision.}
  \label{fig:arima_variants}
\end{figure*}

The broader ARIMA comparison in Fig.~\ref{fig:arima_variants} provides context for the controlled models. It shows that the strongest low-order specifications separate from several higher-order alternatives over the evaluation horizon. This is descriptive evidence from one season rather than a basis for claiming universal superiority of a particular ARIMA order.

Sequential updating does not automatically dominate a static lineup. It changes only within-squad deployment, and its effect depends on forecast updates and realized appearances: for example, ARIMA(1,0,1) gains 24 points, Weighted average gains seven, and ARIMA(1,0,0) loses two. This finding reinforces the need to separate lineup adaptivity from transfers or dynamic captaincy.

\subsection{Budget allocation and the starter--bench frontier}\label{sec:joint_results}

\begin{figure*}[!htb]
  \centering
  \includegraphics[width=\textwidth]{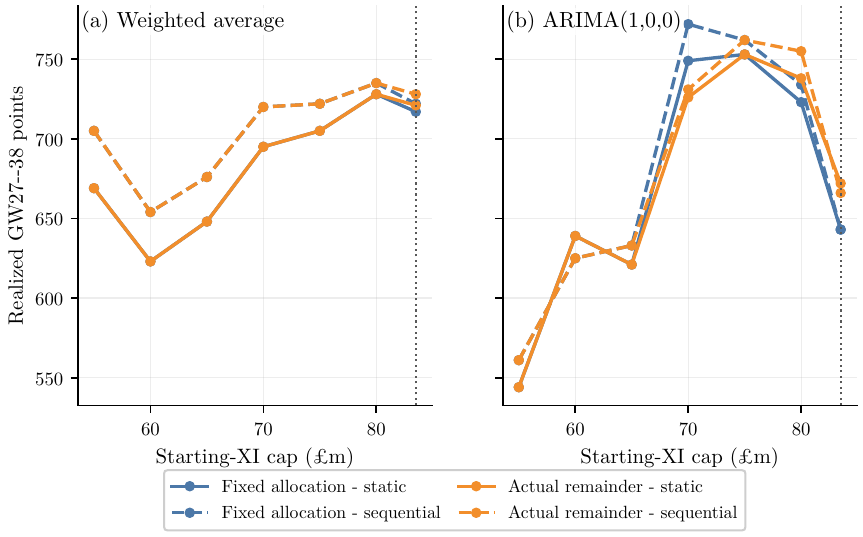}
  \caption{Controlled sensitivity of realized GW27--38 points to the starting-XI cap and bench-budget rule. The vertical line marks the heuristic \pounds 83.5m cap; curves are reported as sensitivity evidence and are not used for ex post cap selection.}
  \label{fig:budget_policy}
\end{figure*}

\begin{figure*}[!htb]
  \centering
  \includegraphics[width=\textwidth]{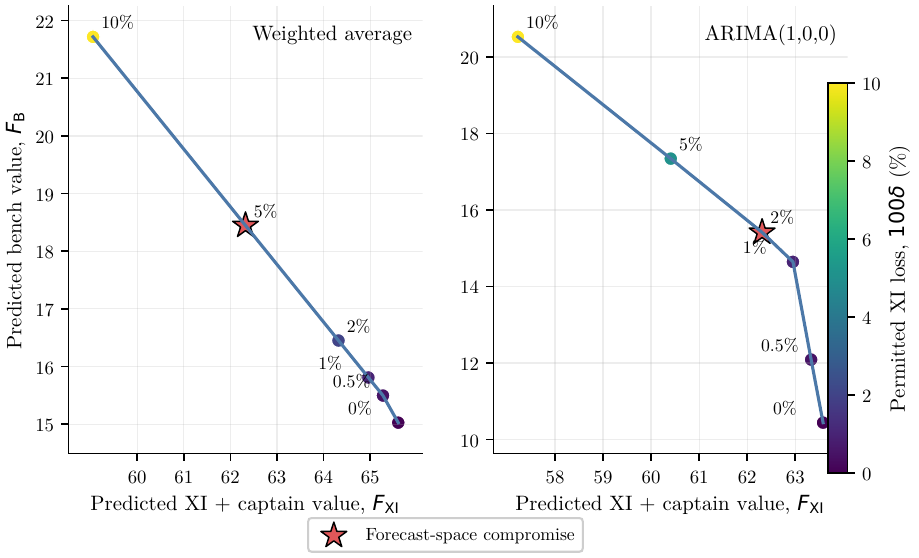}
  \caption{Epsilon-constraint frontiers for predicted starting-XI-plus-captain value $F_{\mathrm{XI}}$ and predicted bench value $F_{\mathrm B}$. Stars mark normalized-distance compromises selected without realized GW27--38 results.}
  \label{fig:pareto}
\end{figure*}

Figure~\ref{fig:budget_policy} confirms that realized performance can be highly sensitive to $b$ and to whether unused XI allowance reaches the bench. Realized-score maxima are therefore not used to select a starting-XI cap. The joint formulation removes $b$, and the forecast-space frontier in Fig.~\ref{fig:pareto} reports the opportunity cost of bench quality.

For ARIMA(1,0,0), the forecast-space compromise permits a 2.0\% XI loss, increasing predicted bench value from 10.439 to 15.414 (47.7\%) and shifting expenditure from \pounds 83.7m/\pounds 16.3m to \pounds 78.1m/\pounds 21.5m. For Weighted average, the selected 5.0\% compromise increases predicted bench value from 15.026 to 18.446 (22.8\%) and shifts expenditure from \pounds 80.2m/\pounds 19.4m to \pounds 72.4m/\pounds 24.2m. All six points for each displayed forecaster are nondominated.

\begin{table}[!htb]
\centering
\caption{Controlled roster-design comparison. Predicted objective values and
squad expenditures refer to the GW27 roster decision; realized points cover
GWs~27--38. The compromise is selected entirely in forecast space.}
\label{tab:joint_comparison}

\setlength{\tabcolsep}{3.5pt}

\begin{tabular}{@{}lrrrrrr@{}}
\toprule
\multicolumn{7}{c}{\textbf{(a) ARIMA(1,0,0)}}\\
\cmidrule(lr){1-7}
Design
& $F_{\mathrm{XI}}$
& $F_{\mathrm B}$
& \shortstack{XI\\(\pounds m)}
& \shortstack{Bench\\(\pounds m)}
& Static
& Sequential \\
\midrule
Fixed allocation, $b=83.5$
& 63.329 & 11.753 & 83.4 & 16.5 & 643 & 643 \\
Actual remainder, $b=83.5$
& 63.329 & 12.087 & 83.4 & 16.6 & 672 & 666 \\
Joint XI-optimal, $\delta=0$
& 63.578 & 10.439 & 83.7 & 16.3 & 743 & 741 \\
Joint compromise, $\delta=2\%$
& 62.309 & 15.414 & 78.1 & 21.5 & 706 & 708 \\
\bottomrule
\end{tabular}

\vspace{1em}

\begin{tabular}{@{}lrrrrrr@{}}
\toprule
\multicolumn{7}{c}{\textbf{(b) Weighted average}}\\
\cmidrule(lr){1-7}
Design
& $F_{\mathrm{XI}}$
& $F_{\mathrm B}$
& \shortstack{XI\\(\pounds m)}
& \shortstack{Bench\\(\pounds m)}
& Static
& Sequential \\
\midrule
Fixed allocation, $b=83.5$
& 65.603 & 10.880 & 80.2 & 16.5 & 717 & 722 \\
Actual remainder, $b=83.5$
& 65.603 & 15.026 & 80.2 & 19.4 & 721 & 728 \\
Joint XI-optimal, $\delta=0$
& 65.603 & 15.026 & 80.2 & 19.4 & 721 & 728 \\
Joint compromise, $\delta=5\%$
& 62.325 & 18.446 & 72.4 & 24.2 & 675 & 722 \\
\bottomrule
\end{tabular}
\end{table}

Table~\ref{tab:joint_comparison} demonstrates two distinct defects of the fixed split. Under ARIMA(1,0,0), the joint XI spends \pounds 83.7m and is infeasible under the heuristic \pounds 83.5m cap; joint selection raises realized points materially. Under Weighted average, the XI optimum spends only \pounds 80.2m, but the fixed rule withholds the unspent allowance and leaves total expenditure at \pounds 96.7m. Passing the actual remainder to the bench makes the two-stage and joint $\delta=0$ solutions coincide. Joint optimization therefore removes an arbitrary cap and can find a globally better roster, but it does not necessarily dominate the actual-remainder two-stage model for every cost vector.

Greater predicted bench value also need not raise realized points over 12 GWs: the compromises score 708 sequential points for ARIMA(1,0,0) and 722 for Weighted average, below the corresponding $\delta=0$ totals of 741 and 728. Bench value is insurance whose realized benefit depends on which starters fail to appear and whether the substitute is formation-valid.

\subsection{External benchmark}\label{sec:benchmark}

\begin{figure*}[!htb]
  \centering
  \includegraphics[width=\textwidth]{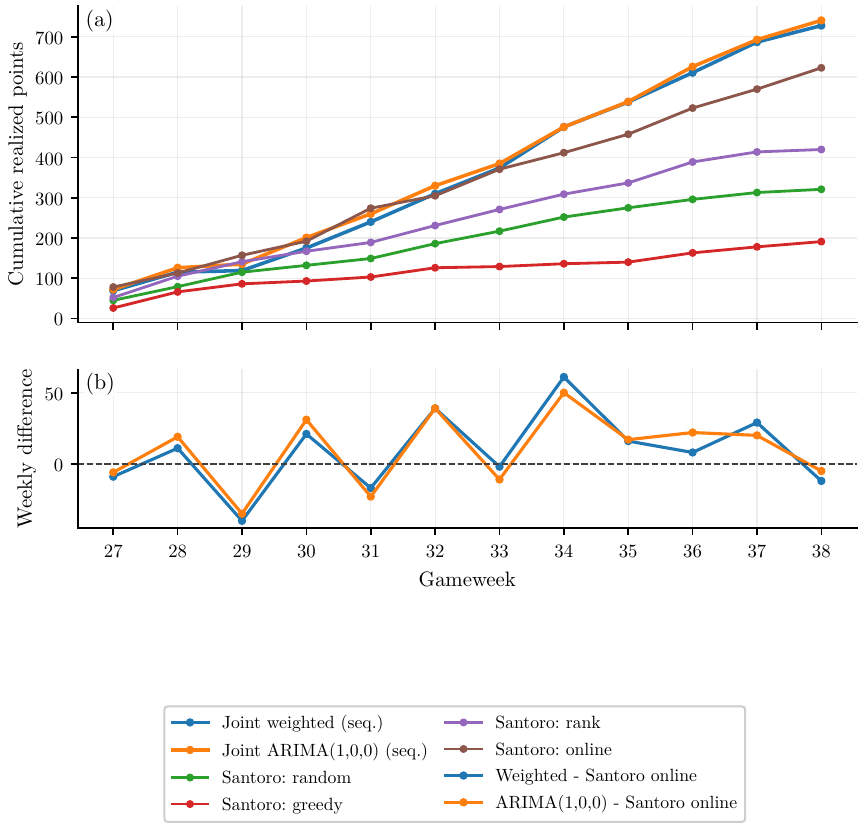}
  \caption{Controlled joint-model comparison with the four externally supplied Santoro strategies over GWs 27--38. Panel (a) gives cumulative points; panel (b) gives weekly differences from the strongest external strategy, mean-MIP-online.}
  \label{fig:external_benchmark}
\end{figure*}

The four benchmark series from Santoro~\cite{santoro2025FPL} were supplied
directly by Santoro for the same GW27--38 interval and are used as provided
external numerical inputs. Their cumulative totals are 191 (greedy), 321 (mean-MIP-random), 420 (mean-MIP-rank), and 623 (mean-MIP-online). Figure~\ref{fig:external_benchmark} compares them with the prespecified joint $\delta=0$ sequential Weighted-average and ARIMA(1,0,0) designs, which score 728 and 741. These baselines are chosen for interpretability and alignment with the structural comparison, not because the test period identified them as best.

The comparison supports competitiveness against these four reported strategies, but not dominance over all FPL decision-support methods. The external series are not recomputed from our raw panel, and implementation details, initial-squad construction, and data snapshots cannot be controlled as tightly as in an internal experiment. We therefore report provenance explicitly and treat the benchmark as descriptive corroboration.

\subsection{Starting-availability sensitivity}\label{sec:availability_results}

\begin{figure*}[!htb]
  \centering
  \includegraphics[width=\textwidth]{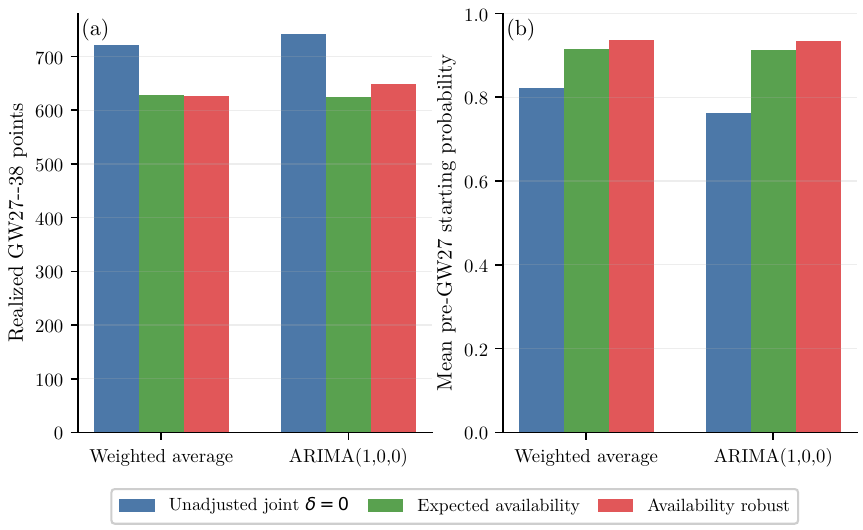}
  \caption{Starting-availability sensitivity for the two representative forecasting methods shown. Complete five-method results are reported in Supplementary Table~S3. Expected-availability and uncertainty-penalized coefficients reduce selection of low-start-probability players, but realized points need not increase.}
  \label{fig:availability}
\end{figure*}

Figure~\ref{fig:availability} illustrates the availability sensitivity for Weighted average and ARIMA(1,0,0). Across the complete five-method results in Supplementary Table~S3, the $\Gamma=0.5$ robust design selects no player with estimated start probability below 70\%; mean squad start probabilities range from 0.926 to 0.936 and minimum probabilities are 0.857. For comparison, the unadjusted ARIMA(1,0,0) roster contains five players below 70\%, has mean probability 0.762, and a minimum of 0.214.

Realized points are mixed. The Simple-average expected-availability roster improves from 625 to 696 static points, whereas Weighted average falls from 721 to 628 and ARIMA(1,0,0) falls from 743 to 625; the corresponding robust totals are 638, 627, and 649. Availability adjustment therefore reduces selection of high-upside players with low estimated starting probabilities, but it is not a performance guarantee. Historical starts may lag injuries, returns, tactical changes, and fixture-specific rotation, and conservatism can remove players who later appear and score.

\section{Limitations}\label{sec:limitations}

The evaluation begins by constructing a new GW27 squad; it does not inherit a manager's GW26 squad or charge for that initialization. The squad is then fixed, so no external transfers, transfer-hit costs, chips, or price-evolution decisions are modeled. The findings concern one-time roster construction and within-squad lineup decisions, not season-long transfer planning.

The empirical evidence covers one Premier League season and a 12-gameweek evaluation horizon. Realized totals are consequently noisy and do not establish universal ordering among forecasts, regularizers, or roster designs. Budget and family plots in the supplement are sensitivity and post-evaluation descriptions, not a basis for choosing a test-optimal model. Multiple seasons and genuinely prospective evaluation are needed for stronger generalization.

The Pareto bench objective is the sum of predicted reserve values, not the expected formation-dependent value of automatic substitutions. The availability model uses smoothed historical starting frequency rather than injury reports, predicted lineups, conditional minutes, or correlated rotation scenarios. A richer stochastic or distributionally robust model could combine appearance probabilities, conditional points, legal substitution paths, and calibrated scenario dependence.

Finally, the Santoro scores are externally supplied. Although they cover the same GWs and provide the most directly relevant integrated FPL comparison available to us, their full data and initialization pipeline is not reproduced inside our code. The external comparison is therefore informative but less controlled than comparisons among our internally recomputed methods.

\section{Conclusion}\label{sec:conclusion}

This study connects player-level forecasting to constrained FPL squad and lineup decisions. The joint MILP makes starter--bench budget allocation endogenous, while the epsilon-constraint frontier makes the cost of stronger bench insurance explicit. In the representative ARIMA(1,0,0) case, joint selection materially improves on both the fixed-allocation and actual-remainder two-stage designs; with Weighted-average costs, joint selection coincides with the actual-remainder rule. The \pounds 83.5m/\pounds 16.5m split therefore serves as a transparent heuristic baseline rather than a general optimum.

The regularization experiments show that dropping $\xG$ and $\xA$, or retaining $\xGI$ as the sole attacking feature, preserves Ridge performance and stability in this sample. LASSO creates a sparser and more stable squad-level decision but sacrifices MAE and realized points. Availability-aware coefficients reduce selections with low estimated starting
probabilities, yet their realized benefit depends on calibration and the cost of excessive conservatism.

Low-order ARIMA and recency-weighted forecasts remain strong, interpretable baselines, but no forecast or policy is uniformly best. The practical value of the framework lies in making each assumption---forecast, uncertainty adjustment, budget allocation, and lineup policy---explicit and auditable. Future work should start from inherited squads and incorporate free-transfer states, transfer hits, chips, price changes, calibrated appearance probabilities, and multi-season prospective validation.

\section*{Acknowledgments}

The authors sincerely thank the reviewers for their constructive comments and valuable suggestions, which have significantly improved the quality and clarity of this manuscript. Tai Dinh is also affiliated with the Kyoto College of Graduate Studies for Informatics (KCGI), Kyoto, Japan. This work was conducted while Tai Dinh was a visiting researcher at Thu Dau Mot University.

\section*{Declaration of competing interest}
On behalf of all authors, the corresponding author states that there is no conflict of interest.

\section*{Author contributions}
Danial Ramezani: conceptualization, data curation, methodology, experiments, visualization, writing, and editing. Tai Dinh: methodology, experiments, visualization, validation, writing, and editing.

\section*{Data availability}
The dataset, source code, computational requirements, and reproduction instructions are available in the project repository: \url{https://github.com/taiduydinh/fplorf}.

\section*{Funding}
Not applicable.

\section*{Ethics approval and consent to participate}
Not applicable; this study uses public sports-performance data and involves no human participants.

\bibliography{_main_v3}

\end{document}